\documentclass{jfm}
\usepackage{graphicx}
\usepackage{epstopdf, epsfig}
\usepackage{subcaption}
\usepackage{amsmath}
\usepackage[normalem]{ulem}
\usepackage[amssymb]{SIunits}
\usepackage{mwe}
\usepackage[section]{placeins}
\usepackage{pdflscape}
\usepackage[export]{adjustbox}
\usepackage{color}

\newcommand{\st}[1]{\makered{\sout{#1}}}

\shorttitle{Enhancing thermal mixing in turbulent bubbly flow by adding salt}
\shortauthor{Pim Waasdorp,  On Yu Dung, Detlef Lohse, Sander G. Huisman}

\title{Enhancing thermal mixing in turbulent bubbly flow by adding salt}

\author{Pim Waasdorp\aff{1}
 , On-Yu Dung \aff{1}
 , Detlef Lohse \aff{1,2},
 \and\\ Sander G. Huisman\aff{1}\corresp{\email{s.g.huisman@utwente.nl}}}

\affiliation{\aff{1}Physics of Fluids Group and Max-Planck Center for Complex Fluid Dynamics, Faculty of Science
and Technology, J.M. Burgers Center for Fluid Dynamics, University of
Twente, 7500 AE Enschede, Netherlands

\aff{2}Max Planck Institute for Dynamics and Self-Organization, Am Faßberg 17, 37077 G\"{o}ttingen, Germany
}

\begin{document}

\maketitle

\begin{abstract}
The presence of bubbles in a turbulent flow changes the flow drastically and enhances the mixing. Adding salt to the bubbly aqueous flow changes the bubble coalescence properties as compared to pure water. Here we provide direct experimental evidence that also the turbulent thermal energy spectra are strongly changed. Experiments were  performed in the Twente Mass and Heat Transfer water tunnel,in which we can measure the thermal spectra in bubbly turbulence in salty water. We find that the mean bubble diameter decreases with increasing concentration of salt (NaCl), due to the inhibition of bubble coalescence. With increasing salinity, the transition frequency from the classical $-5/3$ scaling of the thermal energy spectrum to the bubble induced $-3$ scaling shifts to higher frequencies, thus enhancing the overall thermal energy. We relate this frequency shift to the smaller size of the bubbles for the salty bubbly flow. Finally we measure the heat transport in the bubbly flow, and show how it varies with changing void fraction and salinity: Increases in both result into increases in the number of extreme events.\\
\hfill \break
\end{abstract}

\section{Introduction}

Understanding turbulent mixing is relevant for both industrial processes (e.g. chemical reactors, heat exchangers) and naturally occurring processes (e.g. erosion in rivers, merging at estuaries). At high Reynolds numbers the turbulent velocity fluctuations show an approximate $-5/3$ scaling in the inertial range of the energy spectrum \citep[][]{Kolmogorov1941,Obukhov1949Flow}. Mixing properties of high Reynolds number, turbulent, flows have been the subject of theoretical, numerical and experimental studies \citep{Warhaft2000,Dimotakis2005TurbulentMixing,Villermaux2019MixingStirring}. On top of mixing, by turbulence, an extra enhancement of mixing can be achieved by adding bubbles into the flow. Indeed, rising bubbles, either in a water column or water channel with concurrent pressure driven flow, agitate their surroundings \citep[][]{Magnaudet2000,Mazzitelli2003,Riboux2010,Roig2012,Almeras2015,Almeras2017,Almeras2018,Risso2018,Almeras2019MixingTurbulence,Ruiz-Rus2022CoalescenceSwarm}.
Previous studies have shown the emergence of a bubble-induced scaling in the kinetic energy spectrum, with a $-3$ scaling exponent \citep[see for example][and others]{LanceBataille1991,Riboux2010,Riboux2013ATurbulence,Risso2018,Mathai2020BubblyFlows,Dung2023TheTurbulence}, both in experiments and numerical simulations. Note that in simulations a resolved bubble wake is important for the emergence of this scaling. If the bubbles are modelled as point particles, this scaling is not observed \citep[][]{Mazzitelli2009}, whereas experiments with finite sized particles, fixed in the flow, do show this scaling \citep{Amoura2017}.

The properties of the scalar spectrum in a turbulent flow show remarkable similarity to the kinetic energy spectrum. According to the Kolmogorov--Obukhov--Corrsin theory \citep[][]{Kolmogorov1941,Obukhov1949Flow,Corrsin1951}, the scalar spectrum shows a $-5/3$ scaling in the inertial-convective regime, for high P\'{e}clet and Reynolds numbers \citep{Monin1975}. This only holds when the scalar does not induce buoyancy forces in the flow. Once the scalar can no longer be considered as passive, additional effects and different scaling properties may emerge, for example for turbulent Rayleigh-B\'{e}nard convection \citep{Lohse2010Small-scaleConvection}.

Recent work, both experimental \citep{Dung2023TheTurbulence} and DNS \citep{Hidman2022} shows the emergence of a bubble induced $-3$ scaling in the (passive scalar) thermal spectrum too, similar to the kinetic energy spectrum of bubbly flow. 

By dissolving certain salts into bubbly water, the ability for bubbles to coalesce is inhibited, as was shown by \cite{Craig1993}, who also gave an overview of ion-pairs that alter the coalescence properties. A recent study by \cite{Duignan2021TheInhibition} offers a quantifiable explanation for the inhibition of bubble coalescence by certain ion-pair combinations, based on surface potentials and resulting changes of the Gibbs-Marangoni pressure. Salts that are found in the ocean are among the salts that exhibit this effect. Evidence of this effect can be seen in the foams that are present on shorelines where waves break on the beach, and the absence in fresh water creeks or rivers. Even in very turbulent rivers, the bubbles produced by entrainment of air in rapids disappear quickly. In industrial applications froth flotation comes to mind. Here, air is injected into a mixture of mineral ores, other mining waste products, and water. The ore will act as a surfactant and stick to the bubble. Jointly they rise to the air-water interface ("flotation"), where the ore is separated from the unwanted waste products.

In this work we first study the transition from a single phase $-5/3$ scaling in the thermal energy spectrum of the turbulent flow towards the bubble-induced $-3$ scaling. We do this by performing experiments in our vertical water channel setup where we introduce a small lateral temperature gradient, enough for heat transfer to exist but not enough for the thermally induced buoyant forces to be important. We study the influence of the bubble size, controlled through the salt type and concentration, on the transition frequency between these two scalings. Next, we study the local heat transport properties in these flows, by measuring temperature and velocity simultaneously. We are particularly interested in the effect of the bubble size on the local heat transfer in the flow, again controlled by changing the salt concentration. The paper is organized as follows: Section 2 and 3 describe the experimental setup and how we determine the bubble size, respectively. Section 4 presents the scalar spectra. In section 5 we show the individual velocity and temperature statistics before looking at the local transport properties in section. The paper ends with further discussion and conclusions (section 7).

\section{Experimental setup \& procedure}
\begin{figure}
    \centering
    \includegraphics[width=1\columnwidth]{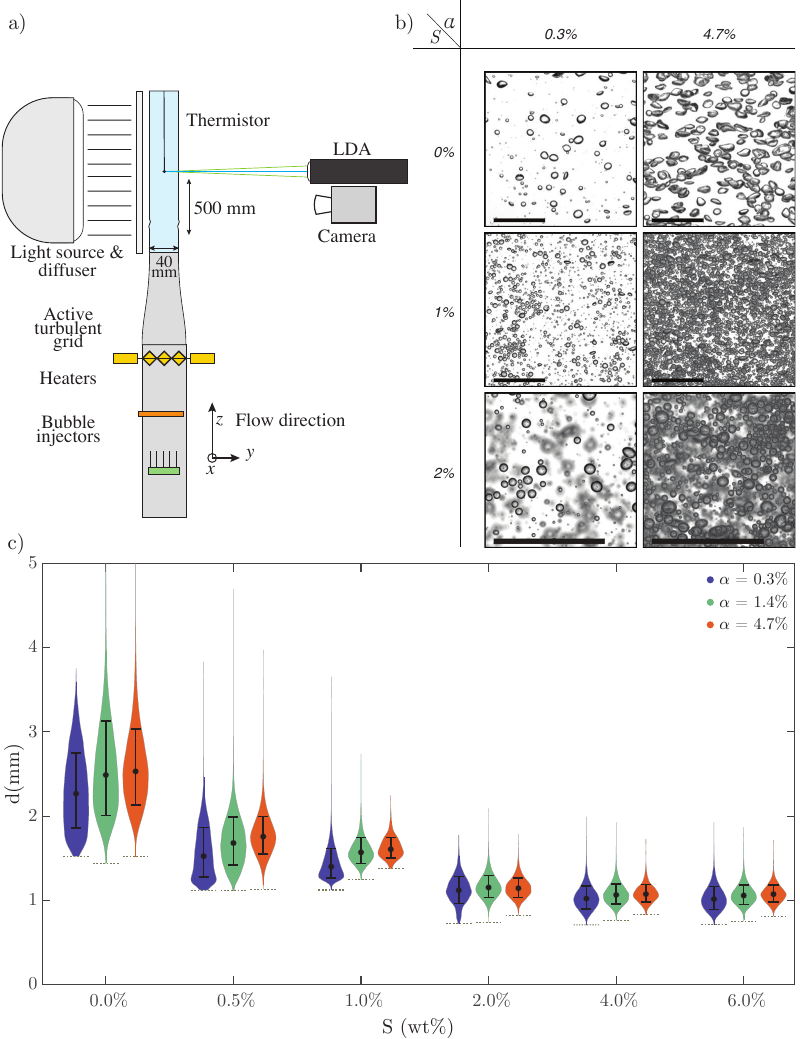}
    \caption{(a) Schematic of the TMHT system and setup of the measurement device. Not to scale. LDA and imaging experiments are not performed simultaneously. The camera is located just in front of the LDA transceiver and is raised before recording to align the field of view of the camera with the measurement location of both the thermistor and LDA. (b) Snapshots of the bubbly flow for low void fraction and high void fraction, and increasing salinity. The length of the scale bar is 20mm. (c) Violin plot of the bubble size distribution as function of the gas volume fraction and salinity. Note that some distributions have an edge at the small diameter end of the violin. This is due to the lower limit (indicated by a small dashed line) of the resolution at which the images were taken and bubble detection was not possible.}
    \label{fig:setup}
\end{figure}

To study turbulent bubbly flow, we use the Twente Mass and Heat Transfer water channel (TMHT), a recently developed and built vertical multiphase water tunnel for studying a variety of turbulent flows \citep[][]{Gvozdic2019b}.  A schematic overview of the tunnel can be seen in figure \ref{fig:setup}a. The driving velocity of the fluid was set to 0.5 m/s, implying a Reynolds number of $\Rey = 2\cdot 10^4$. Six of the twelve heating rods were utilized, all located on the left side of the tunnel, in $x$-direction. The heating rods were powered in pairs by 3 power supply units of 750W each. The active turbulent grid was driven by 15 motors using an algorithm to generate isotropic turbulence \citep[][]{Poorte2002}, although turbulence created in our case is not perfectly isotropic \citep{Dung2023TheTurbulence}. Bubbles were produced by injecting air through a needle bed with 5 transverse rows with 28 needles each, adding up to a total of 140 capillary needles. Air was injected at several different flow rates to generate different void fractions ($\alpha$) of bubbles in the water, in a range of $0$--$4.7\%$. We changed the mean bubble diameter by adding Sodium Chloride (NaCl), in a salinity range of $0$--$6\%\;wt$ (expressed in weight percentage) to our system, which restricts the coalescence of bubbles \citep[see][]{Craig1993} after they have passed through the turbulent grid. Restriction of coalescence results in a lower average bubble diameter, compared to a system without the presence of coalescence restricting electrolytes. 

High speed imaging techniques were used to study the bubbles. The images were taken with a Photron Mini AX200 at a framerate of 1 kHz. An LED lamp  with a diffuser was used for backlighting. Images were recorded for 5 seconds, resulting in 5001 images for each set with the aforementioned framerate. The imaging setup is schematically shown in figure \ref{fig:setup}a. Images were taken at midheight of the test section. Typical images for several different configurations can be seen in figure \ref{fig:setup}b.
Next we describe how we measured the flow velocity. A common technique to study fluid flow velocities is constant temperature anemometry. The benefit of this technique is the very good temporal resolution and thus the ease of studying structure functions and energy cascades. However, in multiphase flow bubble probe interactions make the application of this technique harder as each phase has a different thermal diffusivity. It is possible to overcome this complication as has been shown in other works \citep[][]{Bruun1996,Rensen2005,Almeras2017,Dung2023TheTurbulence}, however, with the goal of also studying heat transport, which requires temperature and velocity to be measured simultaneously, we chose to use Laser Doppler Anemometry (Dantec Dynamics) to record velocity data for the liquid medium. Any possible problems with unwanted interactions between salt and CTA probe, or probe-probe interactions are avoided with this method. The water was seeded with polyamid particles with a monodisperse size of $5 \mu m$, resulting in a particle Stokes number of $St = \frac{t_0U_0}{l_0} \approx 1.8\cdot 10^{-5}$, where $t_0 = \frac{\rho_pd_p^2}{18\mu_w}$ is the relaxation time of the particle (subscript $p$) in the water (subscript $w$), $U_0$ is the averaged single phase flow velocity, and $l_0$ is a characteristic length scale of the water tunnel; in this case taken to be the length in $y-$direction.

A fast response thermistor (Amphenol Advanced Sensors FP07DB204N) was used to record the instantaneous temperature. The thermistor has a response time of 7 ms and a calibrated precision of 1 mK. Calibration of the thermistor was performed with a water bath (PolyScience PD15R-30) and PT100 temperature sensor and fitting the thermistor response to the PT100 measurement using a first order fit to the Steinhart--Hart equation \citep[][]{Steinhart1968}. The voltage potential of a bridge circuit, in which the thermistor acts as a variable resistance, was measured with a lock-in amplifier (Zurich Instruments \texttt{MFLI}), at a sampling frequency of $13.39$ kHz.

To quantify the change of the mixing behaviour in bubbly turbulent flows, as compared to single phase flow, we measured the turbulent heat transport, i.e. the product of the horizontal velocity fluctuations and the temperature fluctuations $( u_x' T' )$. We are interested in this quantity since this, and the turbulent heat variance transport $\langle u_x'^2 T' \rangle$, appears in the balance equation of the passive scalar variance in incompressible flow \citep[][]{Morel2015,Dung2021}, and accurate measurements could prove useful in modelling this quantity. In an ideal experimental setup the measurements of temperature and velocity would take place at the same time and same location. However, this leads to some practical challenges, as mentioned before. When using LDA, we observed an increased sensitivity of the thermistor probe to bubble-probe interactions: where there was no noticeable change in signal before, there are spikes in the signal when LDA is used simultaneously, when the measurements are done in proximity. As origin of this artefact we identified that when the laser from the LDA is reflected inside of bubbles, and directed at the probe, it heats the probe enough to cause a change in the thermistor signal. This artefact can be avoided by increasing the distance between the measurement location of the thermistor and the LDA. We will use Taylor's hypothesis to translate the acquired signals in time, using the mean liquid velocity, such that they correspond to being acquired at the same location.

While the presence of a large scale temperature gradient leads to a non-isotropic situation \citep{Warhaft2000}, for characterisation of the flow we assume isotropy.  Measurements for characterisation were done in single-phase flow only.  Using CTA, with Taylor's hypothesis, the second order structure function was used to obtain an estimate of the energy dissipation rate ($D_{LL} = C_2(\epsilon r)^{2/3}$). This was then used to get the Kolmogorov length scale ($\eta = (\nu^3/\epsilon)^{1/4} = 0.22\; mm$), viscous dissipation timescale ($\tau_\eta = (\nu/\epsilon)^{1/2} = 49\; ms$) and Taylor microscale ($\lambda_u = \sqrt{15\nu(2k/3)/\epsilon} = 4.9 \;mm$). Using the dissipation rate, the Taylor-Reynolds number was estimated to be $\Rey_{\lambda_u} = (2k/3)^{1/2}\lambda_u/\nu = 130$.  In a similar fashion, the scalar dissipation rate and Taylor microscale were estimated. The resulting P\'{e}clet and scalar P\'{e}clet numbers, based on the velocity and temperature microscales, were $Pe_{\lambda_u} = 870$ and $Pe_{\lambda_\theta} = 520$ respectively. We only give global parameters of the flow here, for more details of the characterisation of the flow we refer to \cite{Dung2021} and \cite{Dung2023TheTurbulence}.

\section{Bubble detection and characterisation}
We characterize the injected bubbles with imaging techniques. The images taken with the high speed camera were subjected to a few post processing steps. First, a background subtraction was performed using images that were taken without the presence of bubbles. Next, a circular Hough transform was applied to detect the bubbles, their locations, and their radii. A particle tracking algorithm was then used to couple and track bubbles through subsequent images. From the bubble locations in subsequent images, the bubble velocity was determined. Figure \ref{fig:circhough} in appendix \ref{app:bubble} shows a typical result of the Hough transform overlaid on the image that was subjected to the transform. 

Figure \ref{fig:setup}c shows the size distribution of the bubbles as function of both void fraction and salinity in a violin plot. The violin shapes correspond to the size distribution of the bubbles. For every value of salinity we plot three different void fractions, $0.3\% \leq \alpha \leq 4.7\%$. For zero salinity, the results agree with previously found results by \cite{Almeras2017}. In particular, we reproduce the slight increase of the mean bubble diameter with increasing void fraction. With increasing salinity however, we see a decrease in bubble diameter, similar to what \cite{Gvozdic2019b} describe. The presence of only a small amount of salt already has an effect on the size of the bubbles. This is attributed to the fact that NaCl prevents the coalescence of bubbles \citep[][]{Craig1993}. The bubbles passing through the active grid are fragmented by the large shear forces, resulting in smaller bubbles than initially created, as also described in \cite{Prakash2016,Almeras2017}. In fresh water ($S = 0\%$) the bubbles collide and coalesce before reaching the measurement location in the test section. However, when NaCl is added, the coalescence is inhibited, resulting in smaller bubble diameters. Values for the mean bubble diameter can be found in table \ref{tab:bubbleProperties_fitResults} and the distribution of the size in figure \ref{fig:setup}c.

\begin{landscape}
\begin{table}
  \begin{center}
  \begin{tabular}{c|c|c|c|c|c|c|c|c|c|c|c|c|c|c}
    \hline
$S \;(\%\; wt)$& $\alpha\; (\%)$ & $\gamma\; (mN/m)$ & $\rho\; (kg/m^3)$ & $\mu\; (mPas)$ & $U'\; (m/s)$ & $\langle d \rangle\; (mm) $& $\langle \chi\rangle$ & $\langle V_\text{bub}\rangle\; (m/s)$ & $Re_\text{bub}$ & $We_\text{bub}$ & $Ca$ & $f_t\; (Hz)$ & $f_L\; (Hz)$ & \\ \hline
    & 0        &        &  &           &  0.038  &  ---           &  --- &      ---          &  ---  &  ---  &  ---  &   ---     &   ---    &  \\
    & 0.6      &        &  &           &  0.061  & $2.5 \pm 0.5$  &  $1.6 \pm 0.6$    & $0.69 \pm 0.12$   & $436$ &  $1.06$     &  0.0095     &   14.3    &  0.91    &  \\
0.0 & 1.4      &   72.3 & 997 & 1.002 &  0.058  & $2.6 \pm 0.6$  &  $1.8 \pm 0.7$    & $0.66 \pm 0.11$   & $374$ &  $0.75$     &  0.0091     &    7.99   &  0.94    &  \\
    & 3.0      &        &  &           &  0.082  & $2.5 \pm 0.4$  &  $1.9 \pm 0.7$    & $0.61 \pm 0.11$   & $250$ &  $0.35$     &  0.0084     &    4.93   &  0.86    &  \\
    & 4.7      &        &  &           &  0.100  & $2.6 \pm 0.4$  &  $2.0 \pm 0.6$    & $0.59 \pm 0.11$   & $214$ &  $0.25$     &  0.0081     &    4.07   &  0.79    &  \\ \hline
    & 0        &        &  &           &  0.033  & ---            &  --- &     ---           &  ---  & ---   &  ---  &   ---     &   ---    &  \\
    & 0.6      &        &  &           &  0.056  & $1.6 \pm 0.3$  &  $1.49 \pm 0.3$   & $0.63 \pm 0.11$   & $169$ &  $0.25$     &  0.0068     &   17.1   &  0.97    &  \\
0.5 & 1.4      &  72.5 & 1001 & 1.012 &  0.074  & $1.7 \pm 0.3$  &  $1.45 \pm 0.3$   & $0.68 \pm 0.09$   & $255$ &  $0.54$     &  0.0061     &   12.5   &  0.91    &  \\
    & 3.0      &        &  &           &  0.101  & $1.8 \pm 0.2$  &  $1.50 \pm 0.3$   & $0.70 \pm 0.06$   & $290$ &  $0.66$     &  0.0056     &    6.01   &  0.97    &  \\
    & 4.7      &        &  &           &  0.120  & $1.8 \pm 0.2$  &  $1.47 \pm 0.3$   & $0.69 \pm 0.05$   & $237$ &  $0.44$     &  0.0054     &    4.82   &  0.96    &  \\ \hline
    & 0        &        &  &           &  0.046  &  ---           & ---  &       ---         &   --- &  ---  &  ---  &   ---     &   ---    &  \\
    & 0.6      &        &  &           &  0.076  & $1.6 \pm 0.2$  &  $1.18 \pm 0.2$   & $0.55 \pm 0.10$   & $ 26$ &  $0.01$     &  0.0067     &   18.2   &  0.95    &  \\
1.0 & 1.4      &  72.7 & 1005 & 1.022 &  0.080  & $1.6 \pm 0.2$  &  $1.20 \pm 0.2$   & $0.59 \pm 0.08$   & $ 58$ &  $0.03$     &  0.0060     &   11.0   &  0.89    &  \\
    & 3.0      &        &  &           &  0.096  & $1.6 \pm 0.1$  &  $1.15 \pm 0.2$   & $0.61 \pm 0.06$   & $ 78$ &  $0.05$     &  0.0079     &    7.44   &  1.05    &  \\
    & 4.7      &        &  &           &  0.106  & $1.6 \pm 0.1$  &  $1.21 \pm 0.2$   & $0.61 \pm 0.06$   & $ 75$ &  $0.05$     &  0.0089     &    6.54   &  1.07    &  \\ \hline
    & 0        &        &  &           &  0.038  &  ---           &  --- &    ---            &  ---  & ---   &  ---  &   ---     &  ---     &  \\
    & 0.6      &        &  &           &  0.062  & $1.1 \pm 0.1$  &  $1.11 \pm 0.2$   & $0.60 \pm 0.08$   & $ 63$ &  $0.05$     &  0.0077     &   18.7   &  0.95    &  \\
2.0 & 1.4      &  72.9 & 1013 & 1.042 &  0.074  & $1.2 \pm 0.1$  &  $1.15 \pm 0.2$   & $0.65 \pm 0.07$   & $114$ &  $0.16$     &  0.0063     &   11.0   &  0.82    &  \\
    & 3.0      &        &  &           &  0.090  & $1.2 \pm 0.1$  &  $1.17 \pm 0.2$   & $0.65 \pm 0.06$   & $119$ &  $0.17$     &  0.0054     &    9.11   &  0.50    &  \\
    & 4.7      &        &  &           &  0.103  & $1.2 \pm 0.1$  &  $1.15 \pm 0.2$   & $0.63 \pm 0.06$   & $112$ &  $0.15$     &  0.0054     &    6.69   &  0.42    &  \\
\hline
  \end{tabular}
  \caption{Experimental parameters as function of void fraction $\alpha$ and salinity S. Surface tension $\gamma$ taken from \cite{Nayar2014}. Viscosity $\mu$ taken as sea water salinity equivalent from \cite{Chen1973}, in mPas. Viscosity for S = 0.5\% interpolated from other values. Saline density values taken from \cite{Millero2009}. Single phase channel velocity and Reynolds number are $U_{flow} = 0.52 m/s$ and $\Rey = 2.2\cdot 10^4$ respectively. Bubble Reynolds numbers are calculated with the velocity difference relative to the channel flow, measured by LDA.}
  \label{tab:bubbleProperties_fitResults}
  \end{center}
\end{table}
\end{landscape}

What about the effect of surface tension of the gas-liquid interface, which is known to change when salt is added? To exclude the possibility of this influencing our results, we will quantify the change in surface tension with increasing salinity. The surface tension of fresh water at $20\degree C$ is $\gamma_w = 72.3 \;mN/m$ and that of seawater ($S = 35.28\; g/kg, \; T = 20.92 \degree C$) is $\gamma_{sw} = 73.48 \;mN/m$. This results in a negligible change in capillary number ($Ca = \frac{\mu U_{\text{bub}}}{\gamma}$). Capillary numbers relevant to the experimental settings can be found in table \ref{tab:bubbleProperties_fitResults}, where it can be seen that the difference between experiments is small. The capillary number represents the ratio between viscous forces and surface tension forces, and in these experiments it is much smaller than one ($\mathcal{O} (10^{-6})$). From \cite{Nayar2014} we find a fit for the surface tension as function of the salinity, which is used to calculate the surface tension coefficients that are found in table \ref{tab:bubbleProperties_fitResults}. The change in capillary number is so small that we conclude that a change in surface tension due to increasing salinity can not be the driving mechanism behind the changing bubble size, but rather the physical destruction by passing through the active grid, and the following restriction in the ability to coalesce, as described by \cite{Duignan2021TheInhibition}.

\section{Temperature spectra}

\begin{figure}
    \centering
    \includegraphics[width=1\columnwidth]{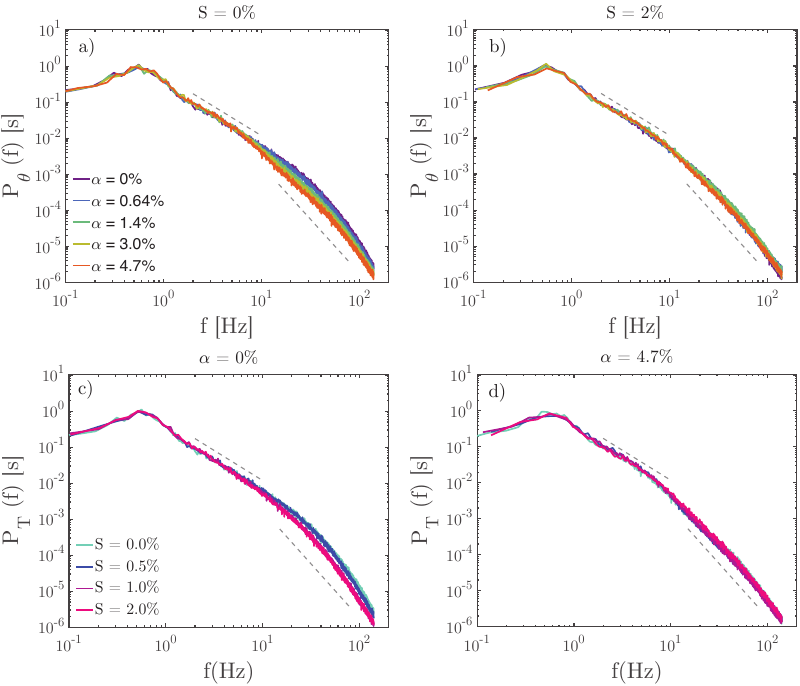}
    \caption{Temperature spectra, normalised by their respective variance, for varying salinity and void fractions. Note that all spectra have a cut-off at 140 Hz, due to the response time of the thermistor. The dashed grey lines in each figure correspond to the classic $-5/3$ scaling and the bubble induced $-3$ scaling. (a) Thermal spectrum as function of $\alpha$, for salinity $S = 0\%$. (b) Thermal spectrum as function of $\alpha$ for high salinity $S = 2\%$. (c) Thermal spectrum as function of salinity for single phase flow. (d) Thermal spectrum as function of salinity for high void fraction $\alpha = 4.7\%$.}
    \label{fig:tempspectra}
\end{figure}

Figure \ref{fig:tempspectra} shows the temperature spectra obtained from measurements with the thermistor. We performed measurements for a duration of 20 minutes. Figure \ref{fig:tempspectra}a shows the spectra as a function of $\alpha$, with a constant and zero salinity. Figure \ref{fig:tempspectra}b shows the spectra for low $\alpha$ as function of salinity. Figure \ref{fig:tempspectra}c shows the spectra for high salinity $S = 2.0\%$ as function of $\alpha$. Finally, figure \ref{fig:tempspectra}d shows the spectra for high $\alpha$ as function of the salinity. The grey dashed lines correspond to the single phase $-5/3$ scaling and the bubbly flow $-3$ scaling. Figure \ref{fig:tempspectra}c, essentially the single phase comparison figure, shows similar results as observed in previous work by \cite{Dung2023TheTurbulence}. Comparing figures \ref{fig:tempspectra}a and \ref{fig:tempspectra}b, we note that in the frequency range between $5$ and $20$ Hz a $-3$ scaling is reached already at lower void fractions for high salinity than for low salinity. 
Figure \ref{fig:tempspectra}d shows that a $-3$ bubble induced scaling exists for all values of salinity at high void fraction.

For different void fractions the frequency at which the transition from a $-5/3$ scaling to a $-3$ scaling occurs is dependent on the void fraction. To investigate this transition frequency, we apply the parametrisation \citep{Dung2023TheTurbulence}:

\begin{align}
    P_\theta (f) = \frac{(f/f_L)^{\zeta_b}}{[1+(f/f_t)^2]^{\frac{\zeta_b-\zeta_a}{2}}}
    \label{eq:batfit}
\end{align}

\begin{figure}
    \centering
    \includegraphics[width=1\columnwidth]{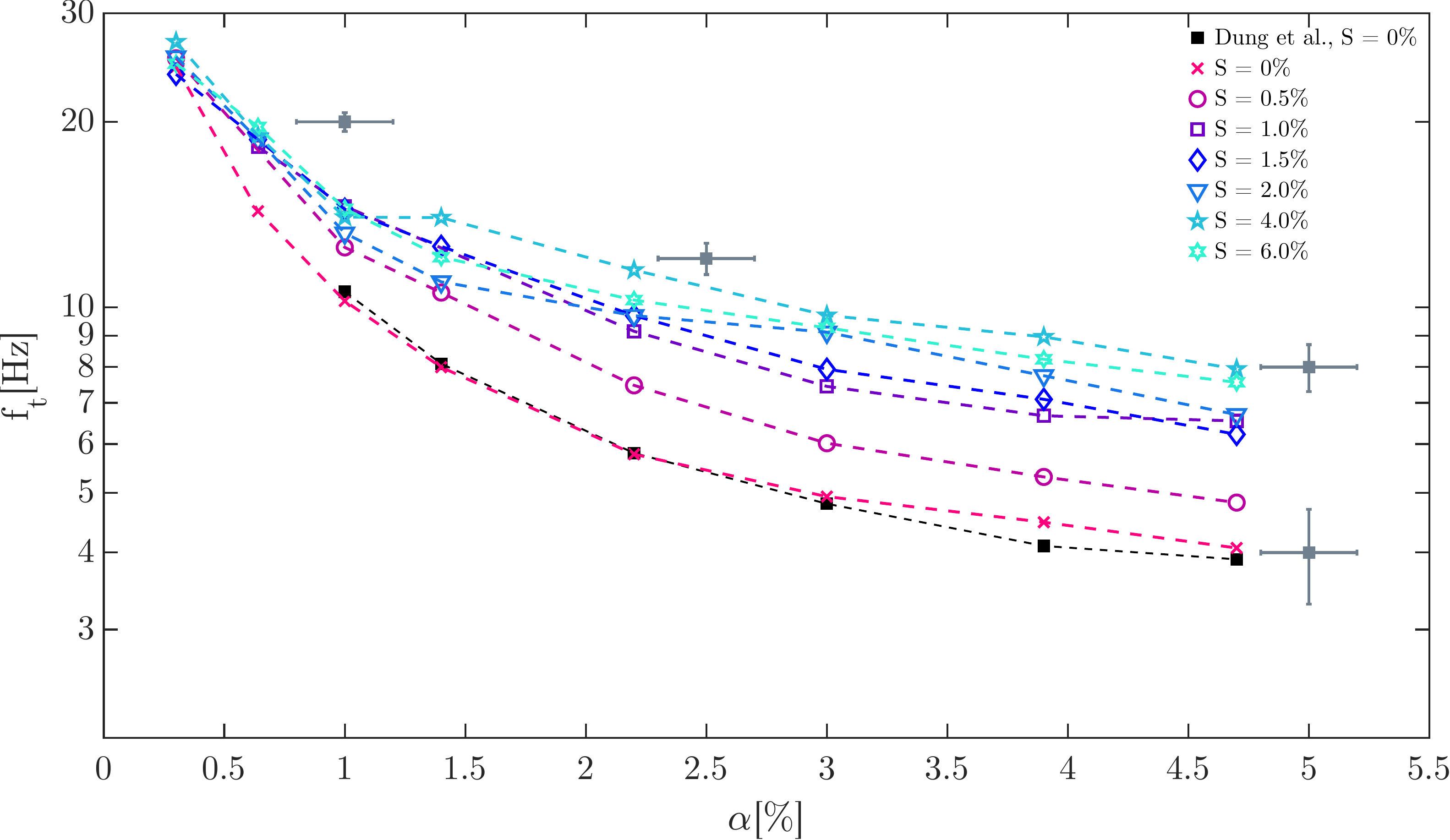}
    \caption{Transition frequency from a $-5/3$ scaling to a $-3$ scaling, as function of void fraction $\alpha$ and salinity $S$. Included are also data from \cite{Dung2023TheTurbulence}, acquired using the same setup, in black squares. The error bars are given on several locations, ensuring good visualization with the log scale on the vertical axis. The error in $f_t$ is determined from the error of the fit. The error in $\alpha$ is determined from the accuracy of the pressure transducer in the setup.}
    \label{fig:transfreq}
\end{figure}

Where $f_L$ and $f_t$ are fitting parameters related to the height of the fit and the transition frequency from the limiting cases $f^{\zeta_b}$ to $f^{\zeta_a}$, where $\zeta_b = -5/3$ (\cite{Kolmogorov1941,Monin1975}) and $\zeta_a = -3$ (\cite{LanceBataille1991,Risso2018}) are the scaling exponents before and after the transition, respectively. The fit is applied in the frequency range from $1$ to $20$ Hz. Figure \ref{fig:batfit} in appendix \ref{app:fit} shows an example of this fit in the transition region for two temperature spectra. This fit was first used in \cite{Batch1951} for the second order structure function, and is the same fit that was used in \cite{Dung2023TheTurbulence}. We apply this fit to all sets of measurements to extract the transition frequency as a function of void fraction and salinity. Figure \ref{fig:transfreq} shows this transition frequency as a function of the void fraction $\alpha$; the color scheme indicates increasing salinity from pink to cyan. Data from \cite{Dung2023TheTurbulence} ($S = 0\%$) are shown in black, and our single-phase results correspond well with these data. There are two notable trends visible in this figure. First, the transition frequency decreases with increasing void fraction. Second, the transition frequency increases with increasing salinity. The appearance of the $-3$ scaling is caused by the presence of the bubbles \citep[][]{Dung2023TheTurbulence,Risso2018} and specifically their wakes since previous numerical work \cite[see e.g.][]{Mazzitelli2009} does not show this effect for point particles. In the frequency range of the $-5/3$ scaling, energy is cascading by momentum transfer from large eddies to smaller eddies, i.e. to larger frequencies. If changing the bubble size influences the spectral power in this region, the typical bubble size should be deducible from the spectra. That is, the eddies produced by the bubbles are related to the bubble size, and influence the thermal and kinetic energy spectra \citep[][]{Dung2023TheTurbulence}. Since the transition frequency and the bubble size change with salinity and void fraction, they surely are connected. Indeed, what we find is that the transition frequency increases with increasing salinity. We relate this to a smaller length scale that triggers this changing scaling, a decrease in bubble diameter (figure \ref{fig:setup}c). We are confident that the bubble size and its resulting wake are affecting the transition frequency. Note that there is a very `steep' region in the low void fraction region. The frequencies in this region can be inaccurate, since a $-3$ scaling region is not fully developed in the energy spectra for these low void fractions. However, it can be seen that for low salinity a change in transition frequency is already apparent.

\section{Individual velocity and temperature distributions}
\begin{figure}
    \centering
    \includegraphics[width=1\columnwidth]{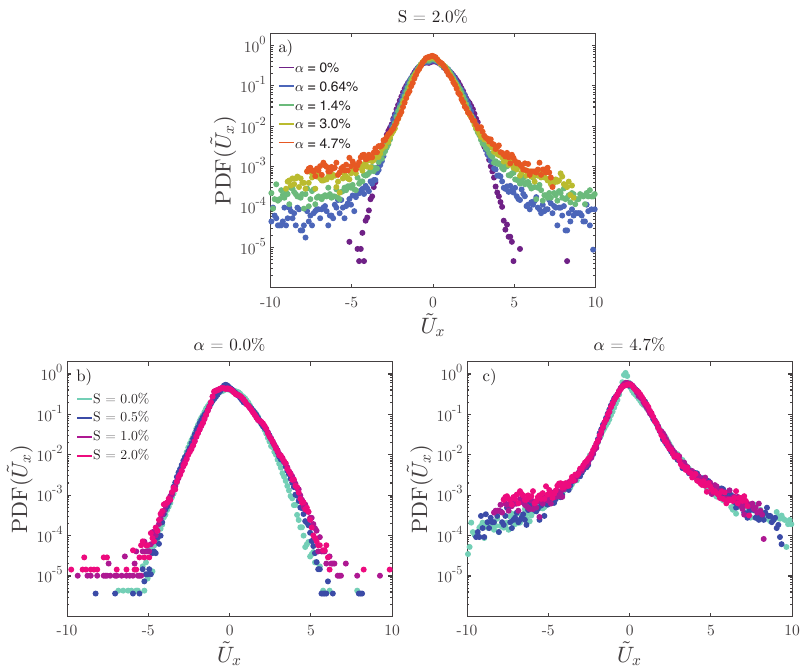}
    \caption{Probability distribution function of the horizontal velocity as a function of void fraction and salinity.}
    \label{fig:velstats}
\end{figure}

Before we look at the turbulent heat transport, we will look at the individual contributions of its components, the velocity and temperature and their distributions. We use LDA to acquire $\mathcal{O}(10^5)$ data points which cover a duration of 20 minutes. Figure \ref{fig:velstats} shows probability distributions of the velocity in horizontal direction. Figure \ref{fig:velstats}a shows the horizontal velocity distribution (standardised as $\tilde{U} = \frac{(U-\langle U\rangle )}{\sigma(U)}$) as a function of the void fraction, for a salinity of $S = 2\%$, and figures \ref{fig:velstats}b and \ref{fig:velstats}c show the horizontal velocity distribution as a function of salinity for single phase and high void fraction, respectively. For the zero salinity results we refer to \cite{Dung2021}. The distributions show agreement with this earlier results for zero salinity. With increasing void fraction the distribution narrows around the mean, and increases in the tails. This increase in the tails highlights the increased occurrence of extreme events \citep[as described by][among others]{Warhaft2000} and mixing in the liquid due to the increasing amount of energy injected by the bubbles. 
In figure \ref{fig:velstats}c we note that a similar increase in intermittency due to the presence of the bubbles is seen. However, this is more pronounced for higher salinity.  This change can be related to the changing bubble size and aspect ratio. 

\begin{figure}
    \centering
    \includegraphics[width=1\textwidth]{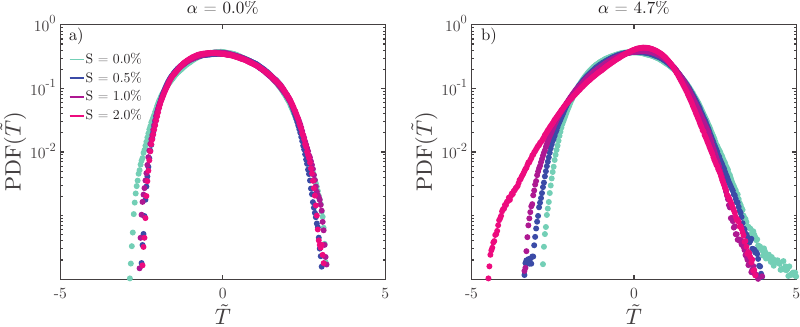}
    \caption{Probability density function of the standardised temperature $\tilde{T} = \frac{T-\langle T\rangle}{\sigma(T)}$, as function of salinity. (a) Single phase temperature distributions. (b) Temperature distributions for $\alpha = 4.7\%$.}
    \label{fig:thermpdfcs000}
\end{figure}

Next we will discuss the temperature distributions. As for velocity, the temperature was measured for both increasing void fraction and increasing salinity. The void fraction was varied from $0$ - $5\%$ and salinity was varied from $0$ - $2.0\%$. Figures \ref{fig:thermpdfcs000}a and \ref{fig:thermpdfcs000}b  show the measured temperature distributions as functions of void fraction and salinity. The temperature here is normalised by subtracting the mean and dividing by the standard deviation, $\tilde{T} = \frac{T-\langle T\rangle}{\sigma(T)}$. Figure \ref{fig:thermpdfcs000}a shows the temperature distribution for the single phase case, for varying salinity. As can be expected this hardly varies, since there are no bubbles. Figure \ref{fig:thermpdfcs000}b shows the PDF of the temperature for high void fraction. Here we see a better defined trend with an increase of the salinity, where the tails get wider with salinity. Note that the high salinity distribution is skewed. We attribute this to the fact that measurements are done in the middle of the test section of our water tunnel setup. As was shown by \cite{Dung2021} (chapter 5), it is possible that the equilibrium location where homogeneity is best shifts with increasing void fraction. The increased skewness for high salinity indicates that the bubble properties attribute to this shift as well.

Consider a very small volume in the liquid that is subjected to a flow with larger bubbles with aspect ratios $\gg 1$, and a flow with small bubbles with aspect ratio $\approx 1$, separately at two different times. One can imagine that the larger bubbles produce larger fluctuations, but less of them, due to their size and the limited inflow of air. For given gas volume, smaller bubbles are larger in number and produce more agitations, even though they might be smaller. This can explain the increased intermittency with increasing salinity as well as increasing void fraction, since for higher $\alpha$ there are more bubbles present creating the agitations. This is indicated by the Weber number decreasing with added salinity as compared to the non-saline conditions. Note that this decrease does not seem to be monotonic with an increase of salinity, but changing velocity, shape, and diameter are also contributing factors.

\begin{figure}
    \centering
    \includegraphics[width=1\columnwidth]{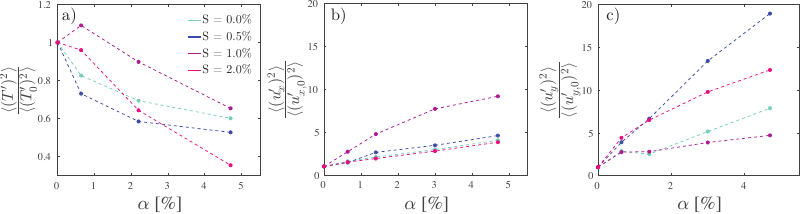}
    \caption{Normalized variance of temperature and velocity as a function of void fraction and salinity.}
    \label{fig:TVVAR}
\end{figure}

Figure \ref{fig:TVVAR} shows the variance of both temperature and velocity (in horizontal and vertical direction) as a function of void fraction and salinity. The variances here are normalised by the single phase variance. Trends that are present for zero salinity water are shown here as well with increasing salinity, but a clear monotonic trend with salinity is absent.

\section{Turbulent heat transport}
\begin{figure}
    \centering
    \includegraphics[width=1\columnwidth]{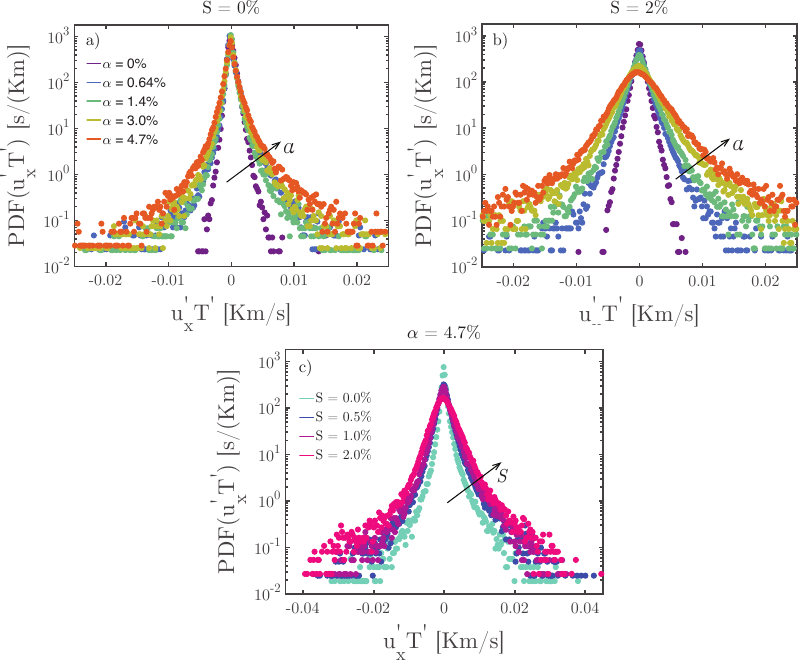}
    \caption{PDFs of the heat flux as a function of void fraction $\alpha$ and salinity $S$. (a) PDFs for varying $\alpha$ for $S = 0\%$. (b) PDFs for varying $\alpha$ for $S = 2\%$. (c) PDFs for varying $S$ for $\alpha = 4.7\%$.}
    \label{fig:fluxpdf}
\end{figure}

\begin{figure}
    \centering
    \includegraphics[width=1\columnwidth]{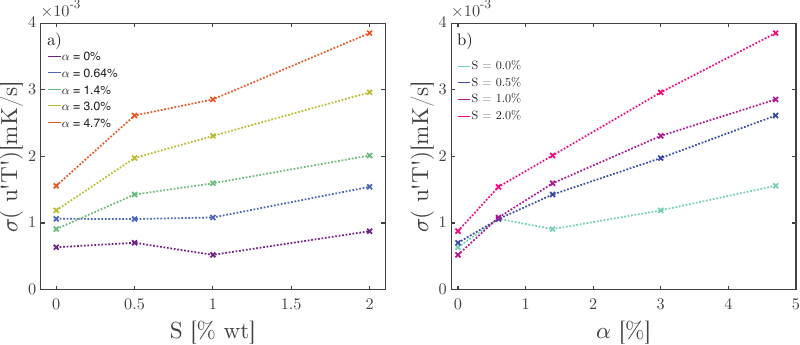}
    \caption{(a) Standard deviation of the heat transport fluctuations as a function of salinity and void fraction. Note that the single phase case is included and stays more or less constant. (b) Standard deviation of the heat transport fluctuations as function of salinity and void fraction, now with $\alpha$ on the horizontal axis.}
    \label{fig:stdflux}
\end{figure}

The efficiency of a process where mixing of a single or multiple scalars is desired, is highly dependent on the scalar flux and its distribution. Figure \ref{fig:fluxpdf} shows the distributions of the measured heat flux. We call the product of the fluctuations of horizontal velocity and temperature heat flux, since this describes a transport of heat. Figure \ref{fig:fluxpdf}a shows the heat flux for zero salinity, as a function of void fraction, where figure \ref{fig:fluxpdf}b shows the heat flux for high salinity. Finally, figure \ref{fig:fluxpdf}c shows the heat flux for high void fraction as a function of salinity. Similar trends as have been discussed are visible here. That is, the tails of the distributions increase both with void fraction and with salinity, indicating that the reducing bubble size, and changing bubble shape increase the intermittency of this property, ultimately increasing the mixing properties of the flow. In the definition of \cite{Villermaux2019MixingStirring}, ``Mixing is stretching-enhanced diffusion...", our results show enhanced mixing of the scalar field (temperature) in our experiments. Recall that the imposed temperature gradient is small such that the temperature acts as a passive scalar, and does not induce any buoyancy driven flow. An increase in intermittency then means that small parcels of relatively hot or cold fluid are increasingly moved around the fluid field. The advective timescales are larger than the diffusive timescales, meaning the movement of these parcels will have a stretching effect on the parcels.  So by stirring the fluid through the forced turbulence from the turbulent grid, and the induced turbulence from the bubbles, the mixing of the scalar field is enhanced. The amount of additional stirring is dependent on the shape and size of the bubbles. Figure \ref{fig:stdflux}a and b show the standard deviations of the measured heat flux signals. It can be seen that for both increasing void fraction and increasing salinity the standard deviation increases, confirming our earlier findings, that mixing is enhanced by adding bubbles of different shapes and sizes. 

To get more insight into the measured heatflux we will look at the passive scalar variance balance equation, for incompressible Navier-Stokes flow \citep{Monin1975,Morel2015}. Using the assumptions that the flow is statistically stationary and unidirectional, the mean temperature gradient is in horizontal direction, and diffusive variance flux can be neglected. Using a boundary layer approximation for the turbulent boundary layer, the passive scalar variance equation can be written as \citep{Monin1975,Libby1975,LaRue1981a,LaRue1981b,Ma1986,Lumley1986EvolutionLayer,Morel2015,Dung2021}:

\begin{equation}\label{eq:balanceEqn2}
 \langle u_z \rangle \frac{\partial \langle T'^2\rangle}{\partial z} =   -2 \langle u_x'T' \rangle \frac{\partial \langle T \rangle}{\partial x}    - \frac{\partial}{\partial x} \langle u_x'T'^2 \rangle - 2 \epsilon_{\theta},
\end{equation}

with the $x$ and $z$ referring to the horizontal and vertical direction, respectively, and the scalar fluctuation dissipation rate $\epsilon_\theta \equiv \kappa\langle (\nabla T')^2\rangle$. Especially interesting is the question whether the heat flux can be accurately modelled. Following \cite{Libby1975, Pope2000,Combest2011OnFlows} and using the eddy diffusivity hypothesis, both the heat flux and variance flux can be written as:
\begin{align}
    -\langle u_x'T' \rangle \equiv \kappa_{t,xx}\frac{\partial \langle T \rangle}{\partial x},\\
    -\langle u_x'T'^2 \rangle \equiv K_{xx} \frac{\partial \langle T'^2 \rangle}{\partial x},
\end{align}
with non-universal diffusivities $\kappa_{t,xx}$ and $K_{xx}$. Accurate measurements can provide information on the modelling of these diffusivities. Figure \ref{fig:fluxpdf}a through \ref{fig:fluxpdf}c show the heat flux as functions of void fraction and salinity, as discussed before. Our measurements however are not appropriate for the approximation of either of the diffusivities mentioned above. The mean value is too close to zero, and there is no correlation between the velocity and temperature measured. 

\section{Conclusions}
In this work we investigated the influence of NaCl on several aspects of a turbulent bubbly water flow with a thermal mixing layer. First, we have looked at the spectrum of a passive scalar (temperature) and how it evolves with increasing void fraction and salinity. We have shown that an increase in void fraction $\alpha$ leads to a decrease in the transition frequency $f_t$ between the $-5/3$ scaling and the $-3$ scaling in the temperature spectra, whereas an increase in salinity leads to an increase in the transition frequency. The changing transition frequency is attributed to the changing bubble properties, while varying void fraction and salinity. Second, we have looked at transverse heat transfer in thermal turbulent bubbly flow. It is found that the mean of the product of velocity and temperature fluctuations does not change, and these quantities are not correlated in the experiments we have performed. We have, however, shown a monotonic increase in the standard deviation of the same quantity, with increasing void fraction and salinity. More specifically, this is only apparent when bubbles are injected. We conclude from this that the presence of bubbles enhances the mixing properties, and that their size is a very relevant parameter for the change in mixing properties.
\hfill\break \hfill\break {\small
\noindent \textbf{CRediT authorship contribution statement} \\
\textbf{Pim Waasdorp:} Conceptualization, Data Curation,  Formal Analysis, Investigation, Methodology,  Resources, Software, Validation, Visualization, Writing - Original Draft. \textbf{On Yu Dung:} Conceptualization, Methodology, Resources, Software, Writing - Review and Editing. \textbf{Sander G. Huisman:} Project Administration, Supervision, Writing - Review and Editing. \textbf{Detlef Lohse:} Funding Acquisition, Supervision, Writing - Review and Editing.
}
\hfill\break \hfill\break {\small
\noindent \textbf{Declaration of Interests.}
The authors declare that they have no known competing financial interests or personal relationships that could have appeared to influence the work reported in this paper.
}
\hfill\break \hfill\break {\small
\noindent \textbf{Acknowledgements.} We thank Gert-Wim Bruggert, Dennis P.M. van Gils, Martin Bos, and Thomas Zijlstra for technical support. We also thank Bo Soolsma for preliminary heat flux measurements and Chao Sun for regular discussions.\\
This work was supported by The Netherlands Center for Multiscale Catalytic Energy Conversion (MCEC), an NWO Gravitation Programme funded by the Ministry of Education, Culture and Science of the government of The Netherlands. 
}

\newpage
\appendix
\begin{figure}
    \centering
    \includegraphics[width=1\columnwidth]{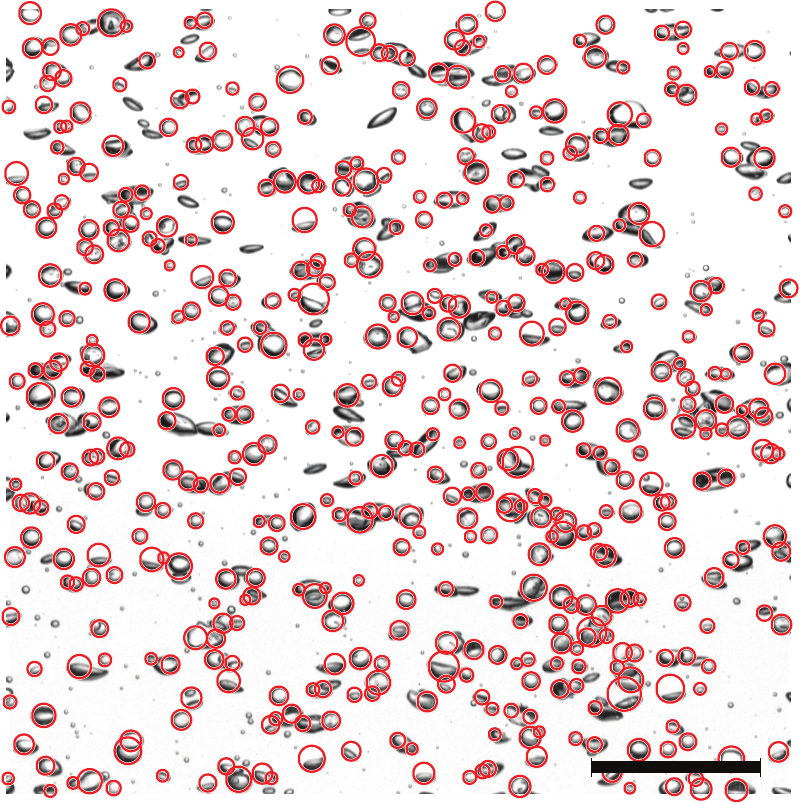}
    \caption{Example image from one of the datasets acquired during the experiments ($\alpha = 1.4\%, S = 0\%$). The red overlaid circles are the result of applying the circular Hough transform. The scale bar has a length of 1 cm.}
    \label{fig:circhough}
\end{figure}
\section{Bubble imaging experiments}\label{app:bubble}
Figure \ref{fig:circhough} shows an example of the result of the circular Hough transform applied to one of our experimental images. Note that this transform is aimed at finding circular objects. This means that there is an inherent error in this detection, since at lower salinities and lower void fractions, a number of bubbles have a non-circular shape. We investigated the aspect ratio of the bubbles manually, and results of this can be seen in table \ref{tab:bubbleProperties_fitResults}. In figure \ref{fig:circhough}, a number of very small bubbles can be seen. The circular Hough transform needs a minimum amount of pixels to be able to identify circles. This is the reason of a sudden end on the lower end of the bubble size distribution in figure \ref{fig:setup}c, especially clear for low void fractions at zero to low salinity. Nevertheless, the trend in bubble size is clear, and not dependent on the detection of an amount of smaller bubbles. We have cross-checked our data by manually fitting ellipses to bubbles, here we find approximately the same bubble size distribution. During the manual investigation of aspect ratios, these smaller bubbles are taken into account.

\begin{figure}
    \centering
    \includegraphics[width=0.85\columnwidth]{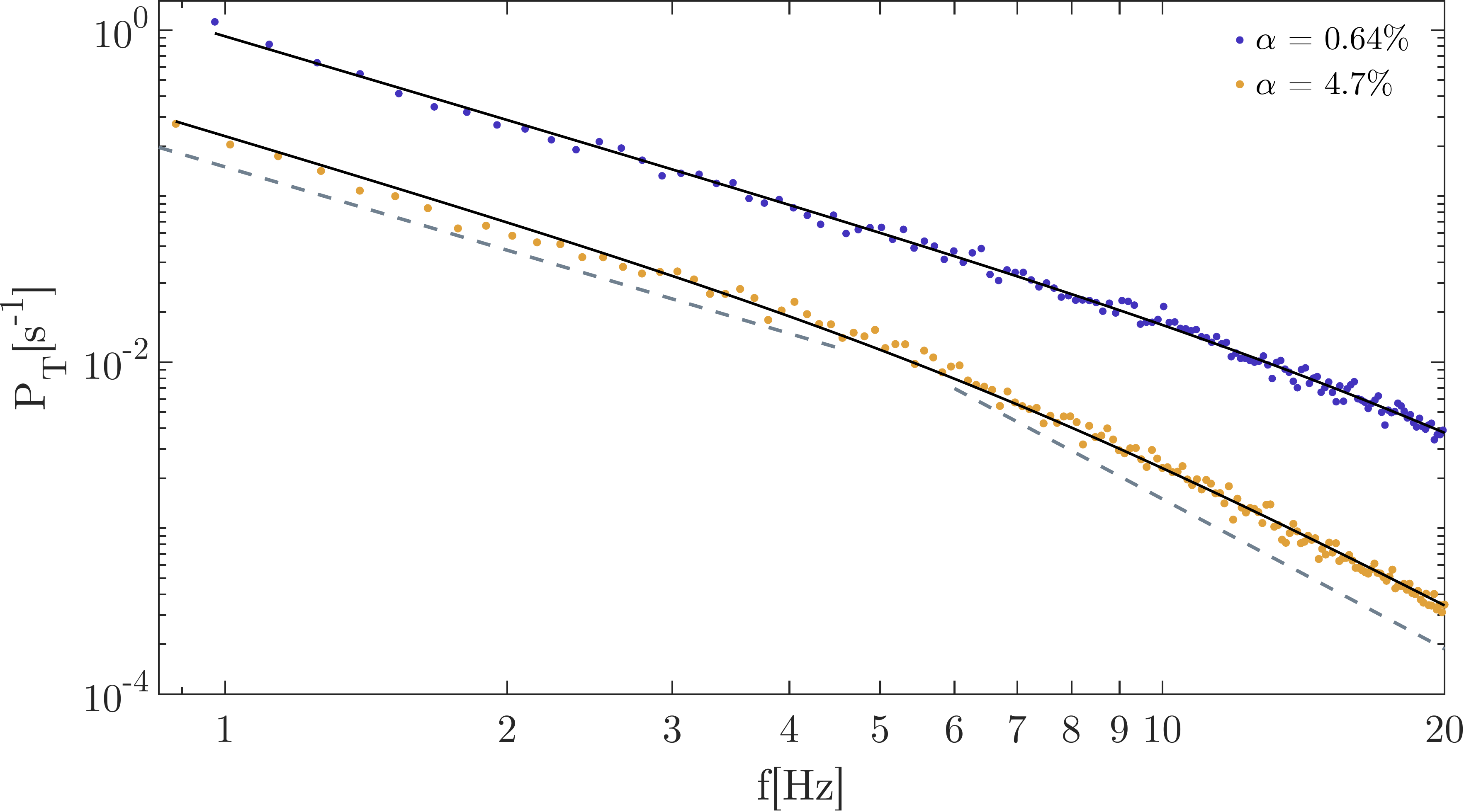}
    \caption{Example of Batchelor fit applied to two temperature spectra in the frequency range where the transition from $-5/3$ to $-3$ happens, for $S = 2\%$. The fit is calculated between $f_z = 1$ Hz and $f_z = 20$ Hz.}
    \label{fig:batfit}
\end{figure}

\section{Accuracy of the fit in the thermal spectra}\label{app:fit}
Figure \ref{fig:batfit} shows an example of two fits as compared to the data they fit, for low and high void fraction and salinity $S = 2\%$. The fit is done using the generalized Batchelor parametrisation shown in equation \ref{eq:batfit}. The grey dashed lines indicate the $-5/3$ and $-3$ scaling from lower to higher frequencies respectively.

\end{document}